# Vortex-glass phase transition and superconductivity in an under- doped (Ba,K)Fe$_2$As$_2$ single crystal

Hyeong-Jin Kim, Yong. Liu, Yoon Seok Oh, Seunghyun Khim, Ingyu Kim, G. R. Stewart+, and Kee Hoon Kim*

CSCMR & FPRD, Department of Physics and Astronomy, Seoul National University, Seoul 151-747, Republic of Korea

## Abstract:

Measurements of magnetotransport and current-voltage ($I$-$V$) characteristics up to 9 T were used to investigate the vortex phase diagram of an under-doped (Ba,K)Fe$_2$As$_2$ single crystal with $T_c$=26.2 K. It is found that the anisotropy ratio of the upper critical field $H_{c2}$ decreases from 4 to 2.8 with decreasing temperature from $T_c$ to 24.8 K. Consistent with the vortex-glass theory, the $I$-$V$ curves measured at $H$=9 T can be well scaled with the vortex-glass transition temperature of $T_g$=20.7 K and critical exponents $z$=4.1 and $\nu$=1. Analyses in different magnetic fields produced almost identical critical exponent values, with some variation in $T_g$, corroborating the existence of the vortex-glass transition in this under-doped (Ba,K)Fe$_2$As$_2$ single crystal up to 9 T. A vortex phase diagram is presented, based on the evolution of $T_g$ and $H_{c2}$ with magnetic field.

PACS numbers:

1. Introduction

The recent discoveries of several families of FeAs-based superconductors such as AFeAsO$_{0.89}$O$_{0.11}$ (A = La, Nd, Sm, and Pr) (FeAs-1111) [1 - 7] and oxygen-free Ba$_{1-x}$K$_x$Fe$_2$As$_2$ (FeAs-122) [8] have triggered a surge of research activities for understanding the basic superconducting mechanism as well as for finding practical applications. These families of compounds  have shown remarkably high superconducting transition temperatures; the highest $T_c$ in the FeAs-1111 system is currently 54 K [1] and in the FeAs-122 system is 37.5 K.[8] In a LaFeAsO$_{0.89}$F$_{0.11}$ polycrystalline specimen, the zero-temperature upper critical field $H_{c2}(0)$ was found to be as high as 65 T.[9] Furthermore, the anisotropy ratio of the upper critical field, i.e., $\gamma = H_{c2}^{\perp c} / H_{c2}^{//c}$ measured in a single crystal of NbFeAsO$_{1-x}$F$_x$ was reported to be about 4 near $T_c$.[10] However, the superconducting gap nature of the FeAs-1111 compounds seems to be yet controversial.[9-11]  Moreover, the investigation of the vortex state is still lacking in the FeAs-1111 series, possibly due to the mostly polycrystalline nature of the samples prepared.

On the other hand, the FeAs-122 compounds exist in the single crystalline form and their transition temperature can be controlled by replacing the alkaline elements (Ba, Ca, and Sr) with K and Na [12 - 15]. Therefore, it is expected that one can investigate anisotropic superconducting properties and vortex phase diagram with variation of $T_c$. In case of an optimally doped Ba$_{0.6}$K$_{0.4}$Fe$_2$As$_2$ single crystal with $T_c \approx 38$ K, $H_{c2}(0)$ along the c-axis was estimated to be 130 T by applying the Werthamer-Helfand-Hohenberg (WHH) formula[16] to the $H_{c2}$ curve measured up to magnetic field $H$=9 T,[17] and the anisotropy ratio $\gamma$ was found to be 3 - 4 near $T_c$.[17 - 19] The vortex solid-liquid phase boundary was suggested from the temperature-dependent irreversibility line in the magnetization curves.[20]  However, these studies have been restricted to the optimally doped samples so far. In addition, the vortex phase boundary based on the irreversibility line needs further investigation of its microscopic transition nature, including the type of phase transition and the critical exponents.

Historically, in order to understand the nature of the vortex phase transition in cuprate superconductors, Fisher et al. proposed[21] a theory of the vortex-glass phase, in which the current-voltage (I-V) characteristics follow a universal scaling function near the vortex glass transition temperature $T_g$. According to this vortex-glass (VG) theory,[21] the vortex-glass correlation length ($\xi$) diverges near $T_g$, described by $\xi \sim |T - T_g|^{-\nu}$ and a correlation time scale $\xi^z$, where $\nu$ is a static exponent and $z$ is a dynamic exponent. The experimental evidence for a vortex-glass phase has been identified for various superconductors such as $YBa_2Cu_3O_7$,[22] $YNi_2B_2C$,[23] and $MgB_2$ [24] through the scaling exponents $\nu$ and $z$ in the range of $\nu \sim 1-2$ and $z > 4$, as predicted by the VG theory.[21] However, to date, the vortex-glass phase transition has not been investigated in detail in the FeAs-based superconductors.

Therefore, the present work reports on the vortex phase diagram and related vortex physics in an under-doped $(Ba,K)Fe_2As_2$ single crystal with $T_c = 26.2$ K, with measurements of magnetotransport, magnetization, and I-V characteristics. We obtained a good scaling of the I-V curves with the critical exponents $z = 4.1$ and $\nu = 1$, uncovering the vortex liquid to glass transition. Using the scaling analyses of the I-V curves and $H_{c2}$ values, we determined the vortex phase diagram of the under-doped $(Ba,K)Fe_2As_2$ single crystal up to 9 T.

## 2. Experiments

The single crystals were grown using the Sn flux-method.[18] The elements Ba, K, Fe, As, and Sn with the molar ratio of 0.6:0.8:2:2:20 were put into an alumina crucible and then sealed inside a quartz ampoule under a partial Ar atmosphere. The ampoule was heated up to 850 $^{o}$C and cooled at 4 $^{o}$C/h. The resultant crystals had a thin plate-like shape. X-ray $\theta$-$2\theta$ diffraction patterns showed that the *c*-axis is parallel to the plate surface. The stoichiometry of the sample was analyzed by using the electron probe micro analysis (EPMA) method, revealing Ba:K=0.64:0.36. Compared with the optimally doped sample $Ba_{0.55}K_{0.45}Fe_2As_2$ with $T_c$~38 K,[8] our crystals corresponds to an under-doped composition.

Two pieces of single crystals were selected with dimensions of 0.056 x 0.7 x 0.7 mm$^3$ and 0.025 x 2.06 x 1.68 mm$^3$ for the investigation of transport properties and magnetization, respectively. Electrical contact was made using silver epoxy (Epotek H20E) and gold wires. We have kept the epoxy curing temperature below 200 $^{o}$C as the samples were found to degrade above this temperature. Temperature-dependent resistivity was measured up to *H*=9 T, both parallel and perpendicular to the *c*-axis, respectively, by using a Physical Property Measurement System (PPMS$^{TM}$, Quantum Design). The *I-V* characteristics were also investigated for various *H* parallel to the *c*-axis by using the PPMS. The magnetization curves were measured up to *H*=9 T, using a vibrating sample magnetometer operating inside the PPMS.

### 3. Results and Discussion

Figure 1 shows the resistivity ($\rho$) curve of the under-doped (Ba,K)Fe$_2$As$_2$ single crystal at zero field. The transition temperature of $T_c$=26.2 K was obtained by choosing the temperature where a linear extrapolation line from the normal state resistivity meets with another line extrapolating the steeply increasing $\rho$ region at the superconducting transition. A superconducting transition-width of $\Delta T_c$~1.2 K was determined from the temperature difference between 90 and 10 % dropoff of $\rho$(30K). The residual resistivity at 30 K of 0.4 m$\Omega$cm was similar to that in Ref. 18, but 10 times higher than that Ref. 20. The residual resistivity ratio is found to be $\rho(300\text{ K})/\rho(30\text{ K})=2.7$, which is lower than that (7.4) of the optimal-doped sample.[17] but is similar to the ratio observed in under-doped material [25].  The inset of Fig. 1 shows the magnetization curve for both zero field cooled (ZFC) and field-cooled (FC) conditions. Assuming the ZFC and FC curves deviate from each other at the superconducting transition temperature, $T_c$ is estimated to be 25.6 K.  This value is close to the value of 26.2 K determined by the resistivity. We also confirmed that the $T_c$ values of several single crystals from the same grown batch were consistent within 0.5 K.

Figure 2(a) shows the magnetic hysteresis curves measured at  temperatures from 8 to 20 K in the field range of $-9$ T$\leq H \leq 9$ T. Even at 20 K, the highest magnetic field in our experiments (9 T) was not sufficient to achieve closure of the magnetic hysteresis curve. Extrapolating these 20 K data, the closure field would be about 10 T, giving at least one value for the irreversibility line (to be discussed below with the phase diagram.)  Figure 2(b) shows the critical current density ($J_c$), estimated from the magnetization curves at various temperatures. The $J_c$ was calculated by using  Bean's critical state model [26] [$J_c=20\Delta M/[w(1-w/3l)]$], where $\Delta M$, $l$ and $w$ are the height of magnetization loops, the length, and the width of the sample ($w<$l), respectively. Under a self field, the value of $J_c$ is found to be ~0.85 MA/cm$^2$ at 4 K. This value is comparable to  the results of the FeAs-122 with $T_c$=36.2 K.[20] When $H$ is increased, the magnetization decreases overall as expected in a typical type-II superconductor above the lower critical field. However, in the data above $T$=10 K, the magnetization curves develop a second peak ($H_{sp}$) as marked with arrows in Fig. 2(a); magnetization increases with field below $H_{sp}$ and decreases above $H_{sp}$. The calculated $J_c$ curve, which is proportional to $\Delta M$, naturally shows a similar peak effect (marked with arrows in Fig. 2(b)). It is interesting to recall that similar phenomena have been also observed in the high-$T_c$ cuprate superconductors. With extensive studies, the origin of the second peak in the cuprates seems to be relatively well understood; for $YBa_2Cu_3O_7$, the second peak effect was attributed to the twin-pinning-center,[27] while for the Bi-based superconductors, it was due to the change of the vortex type from a 3D vortex lattice to a two-dimensional (2D) droplet.[28, 29]

On the other hand, the origin of the second peak in the FeAs system is not clearly understood. The occurrence of the high temperature second peak doesn't seem to be dependent on the material system as it was also observed in the optimal-doped FeAs-122 [20] and $SmFeAsO_{0.89}$ polycrystalline samples.[30] In the former study, the authors suggested that the second peak effect might be due to the vortices pinned by the small sized-normal cores.[20] We note that the optimal-doped FeAs-122 in the former study has been grown with the self-flux FeAs method. What our study shows here is the presence of the second peak even in the under-doped FeAs-122 grown with the Sn flux. As both FeAs-122 crystals show the second peak regardless of the growth flux used, it can be inferred that the normal core, if any, can't come from the remaining impurities from the flux materials in the sample. Therefore, the second peak is likely rather caused by vortex pinning by *intrinsic* defects in the FeAs-122 samples. This observation in turn suggests that the 2nd peak effect in the magnetization loop is rather common in the FeAs superconductors and further studies are required to understand the related vortex pinning mechanism.

To investigate the anisotropic $H_{c2}$ behavior and related vortex physics, we have measured the dc transport properties in magnetic fields applied along and perpendicular to the *c*-axis, as shown in Figure 3. We estimated $H_{c2}(T)$ using the same criterion used to determine the zero field $T_c$ through resistivity: linear extrapolated lines in the normal state and just below $T_c$ were used to find the intersection temperature ($\equiv T_c$). Figure 4 shows the temperature-dependence of $H_{c2}$(T) as determined from the $\rho$(T) curves in Fig. 3. When the slope of $H_{c2}$(T) near $T_c$ is calculated, $-dH_{c2}(T)/dT|_{Tc}$ =3.14 T/K and 6.58 T/K for $H//c$ and $H \perp c$, respectively. The anisotropy ratio $\gamma$ near $T_c$ of 4.2 shown in Fig. 4 is comparable [31] to recent results for (Ba,K)Fe$_2$As$_2$ of $\gamma$~3.5 for a $T_c$ = 30 K sample [19] and $\gamma$~2 for a $T_c$ = 28 K sample [25].

The temperature dependence of $\gamma$ calculated with our data up to 9 T decreases from 4 to 2.7 as T decreases from $T_c$ =26.8 to 24.8 K, as summarized in the inset of Fig. 4. This characteristic decrease of $\gamma$ from ~4 to lower values with the temperature decrease from $T_c$ to lower temperature is consistent with the $\gamma$ behaviors found by several groups in the FeAs-122 systems with $T_c$≈30 K.[17-19,25]

To gain further insight into the vortex-glass physics in our under-doped FeAs-122 single crystal, we measured the *E-J* characteristics up to 9 T (Fig. 5). The *E-J* curves show very similar features to those of YBa$_2$Cu$_3$O$_7$ single crystals and thin films [22] and MgB$_2$ thin film [24] around $T_g$. According to the VG theory,[21] the *I-V* curves in a 3D system show positive curvature for $T>T_g$, negative curvature for $T<T_g$, and a power-law behavior at $T_g$. As shown in Fig. 5, we found that all of these features around $T_g$ exist in our *I-V* results at *H*=9 T with $T_g$=20.7 K with the proviso that only the *low-dissipation regime* (E<4 $10^{-4}$ V/cm) data are considered. As seen in high T$_c$ work (see, e. g., ref. 32), systems quite often show a linear logE vs logI (power law) behavior only below a certain threshold, and this is the case for our data in Fig. 5. Lower dissipation regime data would extend the linear behavior seen at T$_g$ in Fig. 5. Furthermore, the *E-J* curves in the low dissipation regime are well described (Fig. 6) by a universal scaling function with two common variables, $E_{sc} = E/J \, | \, T - T_g \, |^{\nu(z-1)}$ and $J_{sc} = J / |T - T_g|^{2\nu}$ with $T_g$=20.7 K and the critical exponents of $\nu$=1 and $z$=4.1. The critical exponents obtained from the E-J data are indeed in good agreement with the predicted values ($\nu \sim 1$–2 and 4<$z$ ) in the VG theory.[21] When the other *E-J* curves at different magnetic fields (not shown) are analysed, we found that they also follow the same scaling behaviors predicted by VG theory, with the similar critical exponents $\nu$=1 and $z$=4.1±0.2.

It is important to note that in the VG theory, the resistance (see above) should vanish as $(T-T_g)^{\nu/(z-1)}$. Thus, Fig. 7 shows such a plot where indeed the Ohmic resistivity vanishes with the appropriate slope at $T_g$. A further check of the validity of our scaling analysis is whether the glass correlation length (see ref. 33) calculated from our data is with the 3D regime. We arrive at a value of 965 Å, certainly large enough to indicated three dimensional behavior.

Therefore, these results clearly show that a vortex liquid-glass phase transition exists in our under-doped FeAs-122 crystal,the clear evidence of such a vortex liquid-glass transition in the FeAs-122 system.

The inset in Fig. 6 summarizes the vortex phase diagram showing $T_g$ variation with H as obtained from the scaling analyses (not shown) in each field. As discussed above when considering the magnetization data, the 20 K magnetic hysteresis curve data can be extrapolated to meet at about 10 T; this irreversibility point is consistent with the vortex liquid-glass transition line shown in the inset in Fig. 6. This inset also includes low field $H_{c2}$ curve along *H*//*c*. The observation of the vortex-glass phase implies that the FeAs-122 single crystals have strong flux-pinning centers. A priori, either intrinsic or extrinsic defects could be the source for the flux-pinning in the type II superconductors. As for intrinsic defects, the FeAs-122 single crystal could have a structural deficiency, as in the case of the cuprate superconductors forming the collective pinning centers due to oxygen vacancy.[21,22] As for extrinsic defects, the FeAs-122 sample might have small normal cores caused by inclusions of Sn flux (see, e. g., ref. 18) or external ion impurities that can naturally come into the sample during the flux-growth.[18] However, the observation discussed above of the second peak effect in M vs H (Fig. 2a), irrespective of the flux-growth method, suggests that the vortex pinning mechanism in the FeAs-122 single crystals might come from intrinsic defects rather than an extrinsic one such as the Sn flux.

In a recent comparative study of polycrystalline specimens of $NdFeAsO_{0.85}$ and $NdFeAsO_{0.85}F_{0.15}$ [34], the former has shown a clear vortex-glass scaling similar to the observation made here while the latter didn't show any scaling behavior. This observation also supports the idea that intrinsic defects such as oxygen vacancies can play a role in creating the vortex liquid-glass transition in the FeAs based superconductors.

In this regard, more systematic researches are necessary to understand the origin of the vortex liquid-glass transition reported here. For example, synthesis of 122 FeAs superconducting single crystals with systematic control of intrinsic defect levels and studies of the vortex liquid-glass transition would be quite useful.

4. Conclusions

We have measured magnetization, magnetotransport and I-V characteristics in the under-doped FeAs-122 single crystal with $T_c$=26.2 K to study its superconducting properties, upper critical fields, and to investigate the possibility of a vortex glass phase. Analyzing the data using the 3D vortex-glass phase transition theory, we found scaling of the $I$-$V$ curves in each magnetic field up to 9 T, with critical exponents of $\nu$=1 and $z$=4.1, constituting evidence of a vortex glass to vortex liquid transition in the $(Ba,K)Fe_2As_2$ single crystal. The decrease of the upper critical field anisotropy was found to exist with decrease of temperature from $T_c$ to 24.8 K, consistent with the observation made in other FeAs-122 crystals with different $T_c$ values.

Acknowledgements: This work was supported by the National Research Lab program (M10600000238) and KICOS through a grant provided by the MEST (K20702020014-07E0200-01410). Work at Florida supported by the US Department of Energy, contract no. DE-FG02-86ER45268.

\* Corresponding author. Tel.: +82 2 880 9068; fax +82 2 888 0769, E-mail address:

khkim@phya.snu.ac.kr (K. H. Kim)

+ on sabbatical from Department of Physics, University of Florida

FIGURE CAPTIONS

1. The temperature-dependence of resistivity at zero field for the $(Ba,K)Fe_2As_2$ single crystal. The transition temperature is about 26.2 K and the transition width 1.2 K. Inset shows a temperature-dependent magnetization curve measured at $H$=15 Oe applied along the $c$-axis.

2. (a) The magnetic hysteresis curves of the $(Ba,K)Fe_2As_2$ single crystal with $H$//$c$-axis at $T$=8, 10, 15, and 20 K. (b) The critical current density between 4 and 20 K estimated from magnetization curves by using Bean's critical state model. Peaks in both M and $J_c$ vs H are marked with arrows.

3. The temperature-dependence of resistivity along (a) $H$//$c$-axis and (b) $H \perp c$-axis in $H$=0, 0.2, 0.4, 0.6, 0.8 1, 2, 3, 5, 7, and 9 T (from left to right) for the $(Ba,K)Fe_2As_2$ single crystal.

4. The temperature-dependence of the upper critical field $H_{c2}$ for the $(Ba,K)Fe_2As_2$ single crystal. The open and solid symbols represent $H_{c2}$ parallel and perpendicular to the $c$-axis, respectively. Inset: temperature-dependence of anisotropy ratio $\gamma$ determined from the upper critical fields.

5. $E$-$J$ characteristics in the range $T$=18.6–24.0 K with a 0.3 K step under $H$=9 T. The dashed line indicates $T_g$. The negative curvature logE vs logI data expected below $T_g$ can be seen in our data starting at about 19.8 K.

6. The vortex-glass scaling behavior for the low (V<$10^{-5}$ V) dissipation regime . The $I$-$V$ curves collapse very nicely onto two scaling curves, one for above $T_g$ and the other for below $T_g$, near the vortex-glass phase transition temperature of 20.7 K. The inset shows the phase diagram revealing the vortex-glass (VG) to the vortex-liquid (VL) transition.

7. A plot of $(d(\ln\rho)/dT)^{-1}$ vs temperature T shows an intercept near $T_g$ determined from the scaling (Fig. 6), with a slope consistent with the scaling-determined exponents as discussed in the text.

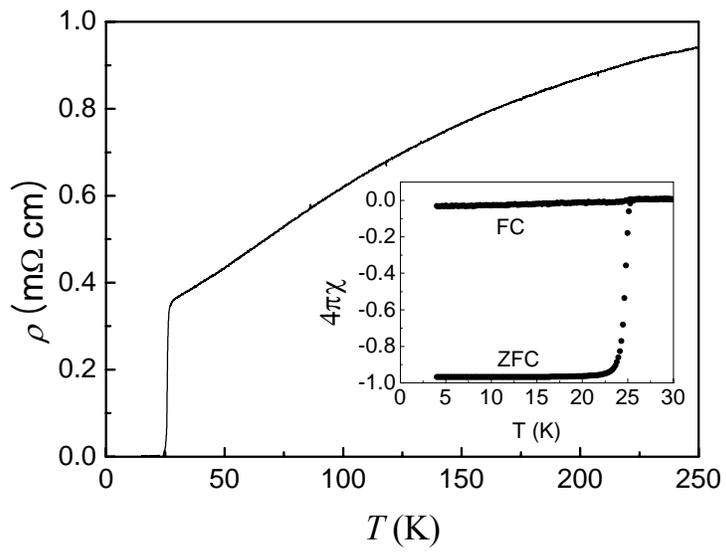

Figure 1. Hyeong-Jin Kim et al.,

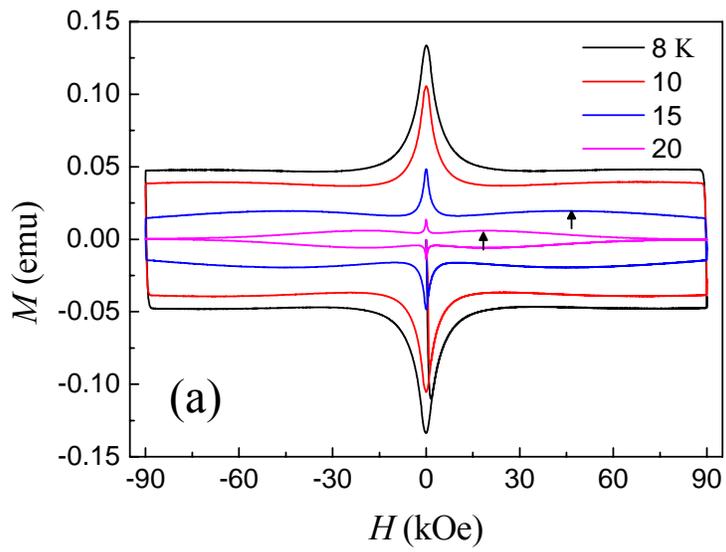

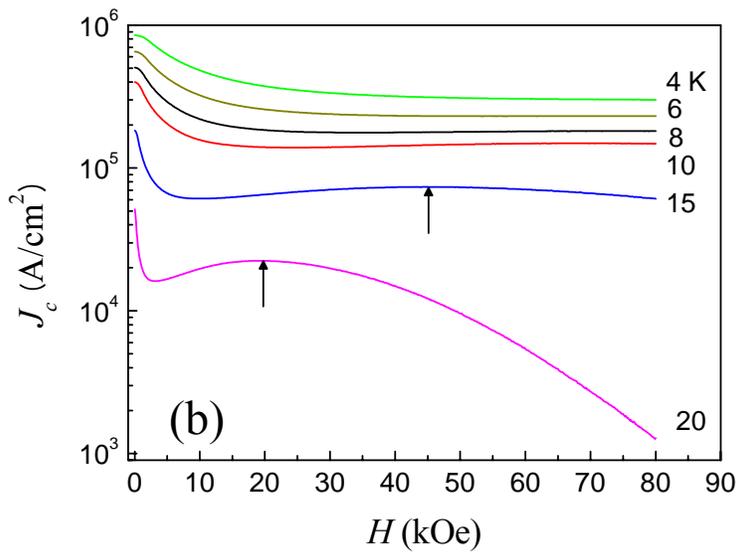

Figure 2. Hyeong-Jin Kim et al.,

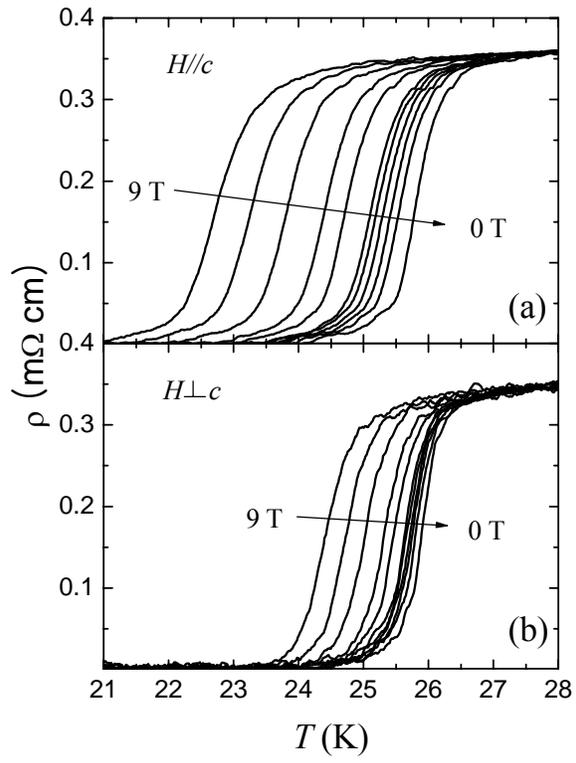

Figure 3. Hyeong-Jin Kim et al.,

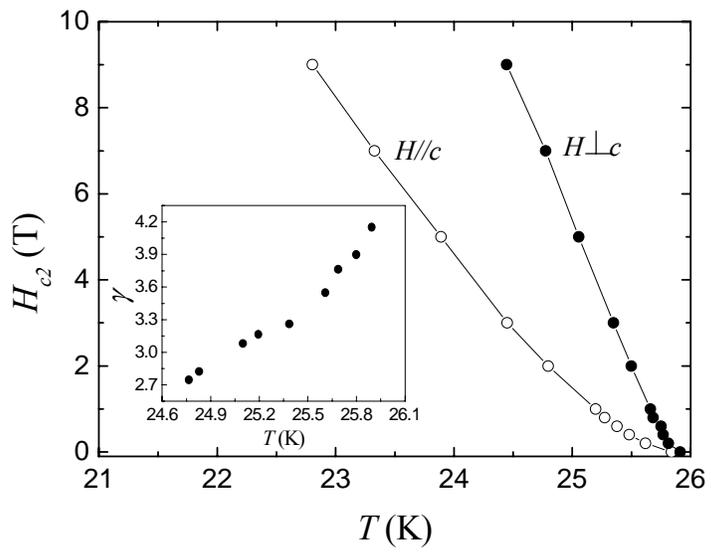

Figure 4. Hyeong-Jin Kim et al.,

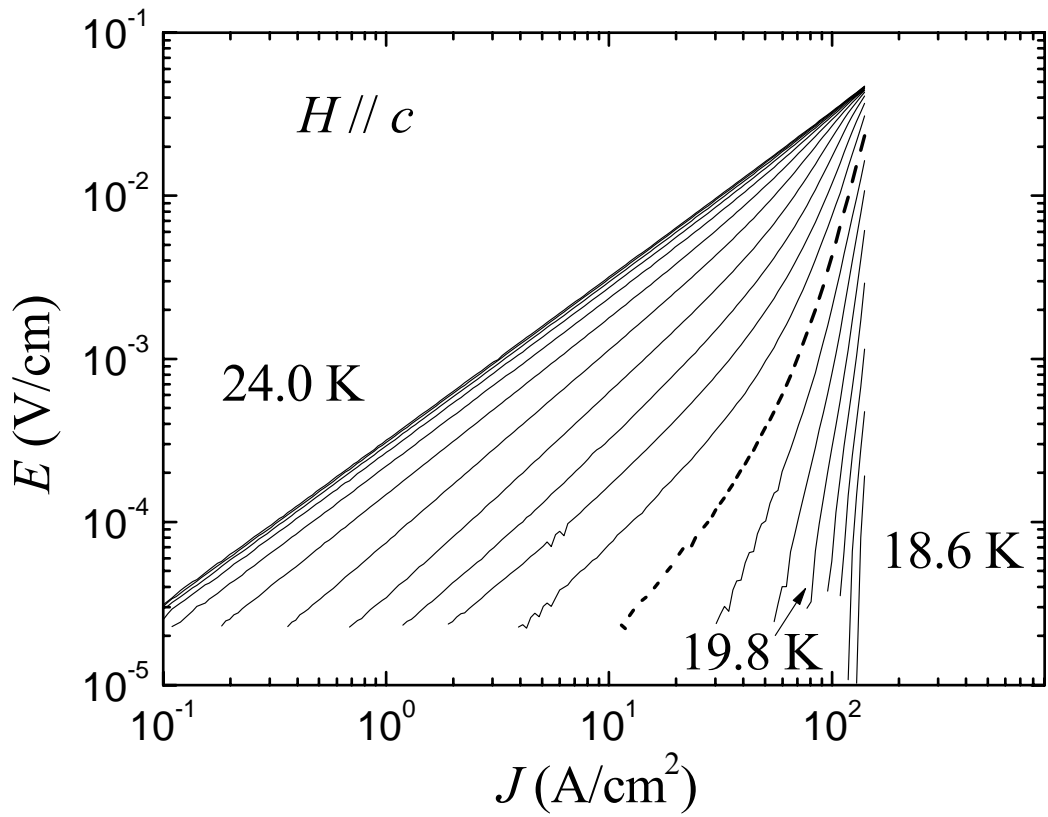

Figure 5. Hyeong-Jin Kim et al.,

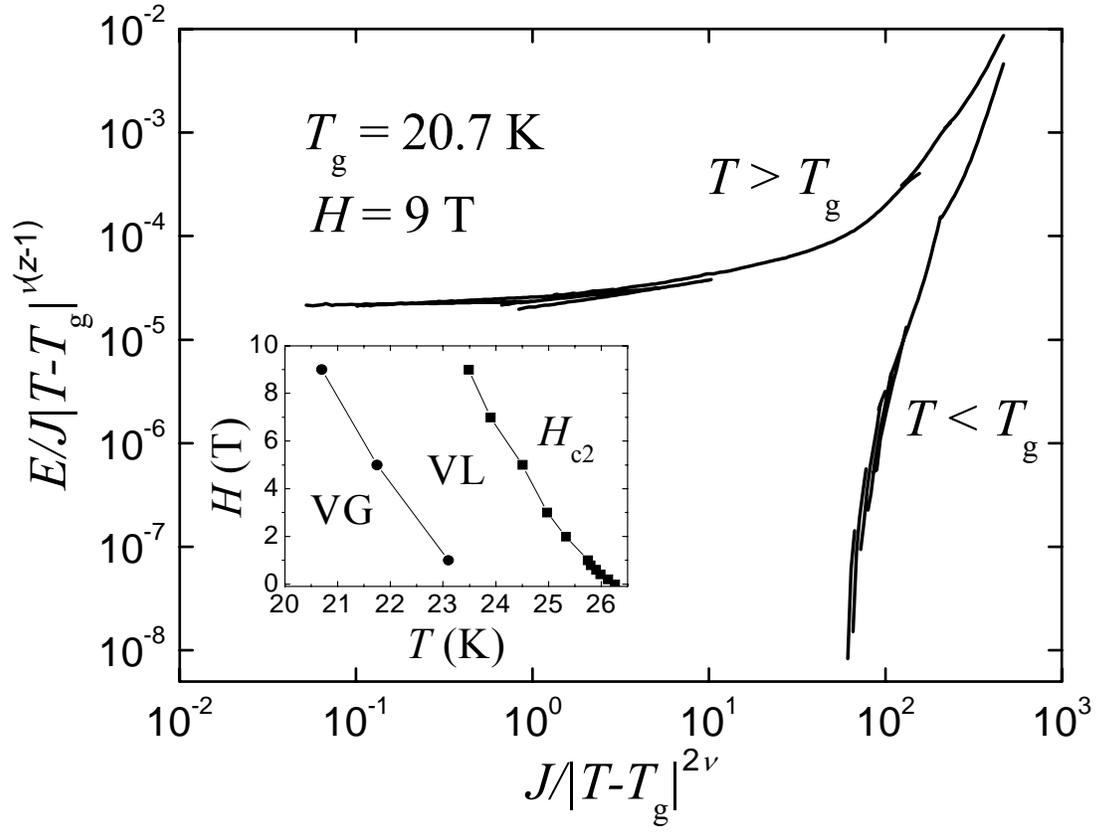

Figure 6. Hyeong-Jin Kim et al.,

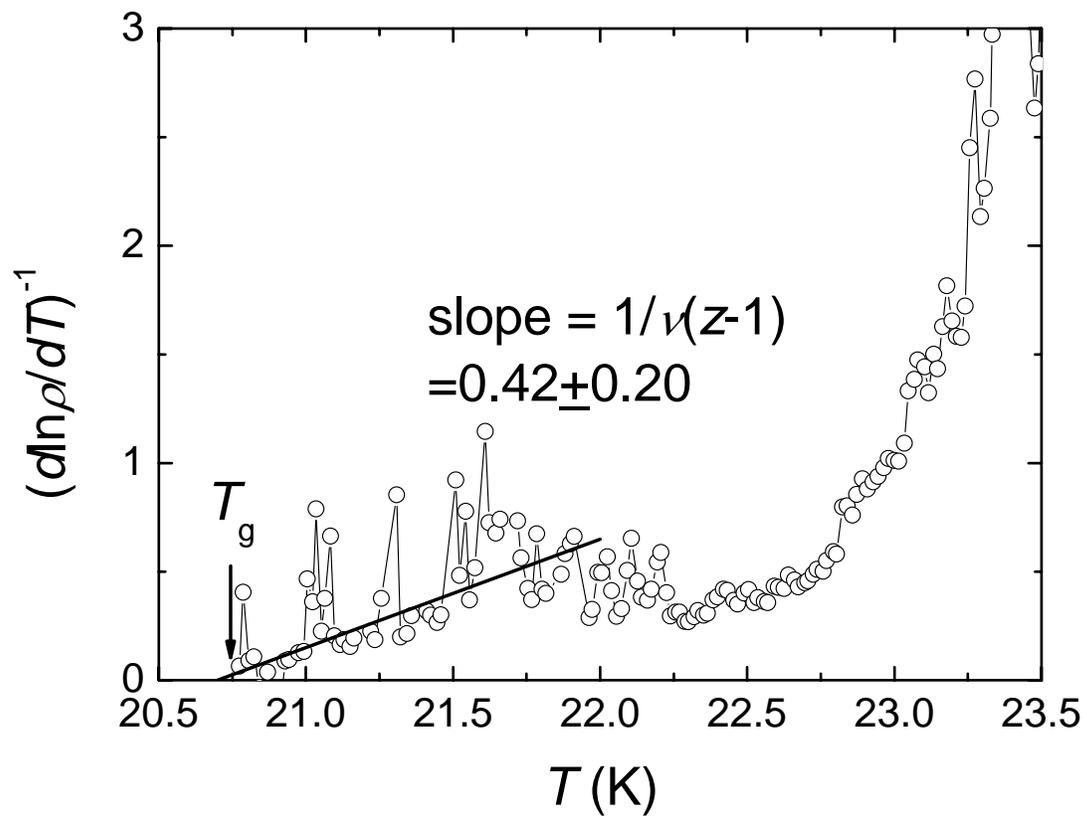

Figure 7. Hyeong-Jin Kim et al.,